\documentclass[12pt,letterpaper,rotate]{article}

\usepackage{amsmath}
\usepackage{amsfonts}
\usepackage{amssymb}
\usepackage{graphicx}
\usepackage[numbers,sort&compress]{natbib}

\usepackage[hang,small,bf]{caption}

\usepackage{rotating}

\DeclareRobustCommand{\SkipTocEntry}[4]{}

\def\beq{\begin{equation}}
\def\eeq{\end{equation}}
\def\bea{\begin{eqnarray}}
\def\eea{\end{eqnarray}}

\def\d{{\rm d}}
\def\l{{\ell}}

\def\edth{\;\raise1.0pt\hbox{$'$}\hskip-6pt\partial\;}
\def\baredth{\;\overline{\raise1.0pt\hbox{$'$}\hskip-6pt
\partial}\;}

\begin{document}

\vspace{5mm} 
\vspace{0.5cm}

\begin{center}
{\Large Causality and Primordial Tensor Modes}
\\[1.0cm]
{ Daniel Baumann$^\dagger$$^{\rm 1, 2}$
and Matias Zaldarriaga$^{\rm 1, 2}$}
\\[0.5cm]
\end{center}

\begin{center}


{\small \textsl{$^{\rm 1}$ Department of Physics, Harvard University, Cambridge, MA 02138, USA}}

{\small \textsl{$^{\rm 2}$ Center for Astrophysics, Harvard University, Cambridge, MA 02138, USA}}

\end{center}

\vspace{2cm} \hrule \vspace{0.3cm}
{\small  \noindent \textbf{Abstract} \\[0.3cm]
\noindent
We introduce the real space correlation function of $B$-mode polarization of the cosmic microwave background (CMB) as a probe of superhorizon tensor perturbations created by inflation.
By causality, any non-inflationary mechanism for gravitational wave production after reheating, like 
global phase transitions or cosmic strings,
must have vanishing correlations for angular separations greater than the angle subtended by the particle horizon at recombination, {\it i.e.}~$\theta \gtrsim 2^\circ$.
Since ordinary $B$-modes are defined non-locally in terms of the Stokes parameters $Q$ and $U$ and therefore don't have to respect causality, special care is taken to define `causal $\tilde B$-modes' for the analysis.
We compute the real space $\tilde B$-mode correlation function for inflation and discuss its detectability on superhorizon scales where it provides an unambiguous test of inflationary gravitational waves.
The correct identification of inflationary tensor modes is crucial since it relates directly to the energy scale of inflation. Wrongly associating tensor modes from causal seeds with inflation would imply an incorrect inference of the energy scale of inflation.
We find that the superhorizon $\tilde B$-mode signal is above cosmic variance for the angular range $2^\circ < \theta < 4^\circ$ and is therefore in principle detectable. 
In practice, the signal will be challenging to measure since it requires accurately resolving the recombination peak of the $B$-mode power spectrum.
However, a future CMB satellite ({\sl CMBPol}), with noise level $\Delta_P \simeq 1\mu$K-arcmin and sufficient resolution to efficiently correct for lensing-induced $B$-modes, should be able to detect the signal at more than 3$\sigma$ if the tensor-to-scalar ratio isn't smaller than $r \simeq 0.01$.

 \vspace{0.5cm}  \hrule
\def\thefootnote{\arabic{footnote}}
\setcounter{footnote}{0}

\vspace{1.0cm}

\vfill 
\noindent
$^\dagger$ {\footnotesize {\tt dbaumann@physics.harvard.edu}}
\hfill \today


\newpage


\section{Introduction}

Inflation  \cite{Guth:1980zm, Linde:1981mu, Albrecht:1982wi}
provides a convincing microscopic explanation for the observed {\it scalar} fluctuations in the temperature of the cosmic microwave background (CMB).
The once popular causal seeds models for structure formation have been ruled out since they fail to reproduce the Doppler peaks of the CMB temperature correlations \cite{Albrecht:1995bg, Hu:1996yt, Allen:1997ag, Pen:1997ae}.
In addition, the {\it superhorizon} nature of the scalar modes produced by inflation leaves a characteristic signature in the temperature-polarization cross-correlation \cite{SZ} that cannot be reproduced by causally-constrained theories.  The dominant mechanism for the generation of scalar perturbations is therefore a period of accelerated expansion 
prior to the hot Big Bang. 
Nevertheless, the constraints coming from the observations of scalar modes ({\it i.e.} temperature and $E$-mode polarization anisotropies and their cross-correlation), don't rule out the possibility that, while subdominant for scalar perturbations, 
causal seeds may be a significant source of {\it tensor} modes. These tensor modes correspond to transverse, traceless metric perturbations or gravitational waves, $g_{ij} = \delta_{ij} + h_{ij}$, where $h^i_i =\partial^i h_{ij} =0$.

$B$-modes of CMB polarization are often described as a `smoking gun' signature of inflationary gravitational waves  \cite{Kamionkowski:1996ks, Zaldarriaga:1996xe}. 
Imagine therefore the glorious day 
that we detect $B$-mode polarization.
How do we prove to the skeptic that the signal really has its primordial origin in inflation?
In this paper we study causality constraints on the $B$-mode correlation function in real space.
This is in the spirit of Ref.~\cite{SZ} who studied the corresponding problem for scalar modes and identified the superhorizon signature in the temperature-polarization cross-correlation function.
Here we extend their treatment to tensor modes and $B$-mode polarization.

\vskip 6pt
Consider concretely two qualitatively different mechanisms for the production of a stochastic background of gravitational waves:
\begin{itemize}
\item First, symmetry breaking phase transitions {\it after} the hot Big Bang ({\it i.e.}~after reheating) may produce a spectrum of tensor modes, {\it e.g.}~\cite{JonesSmith:2007ne}. Furthermore, defect models, {\it e.g.}~\cite{Seljak:1997ii, Pogosian:2007gi}, also create vector and tensor modes that source CMB polarization.
In these models, causality constrains any physical real space correlations to vanish on superhorizon scales. In the power spectrum this corresponds to an approximate white noise spectrum on superhorizon scales.

\item Second, quantum fluctuations during an epoch of inflation {\it before} the hot Big Bang produce a spectrum of long-wavelength gravitational waves. The spectrum is scale-invariant even on superhorizon scales in apparent violation of causality.
\end{itemize}

\begin{figure}[h!]
    \centering
        \includegraphics[width=0.95\textwidth]{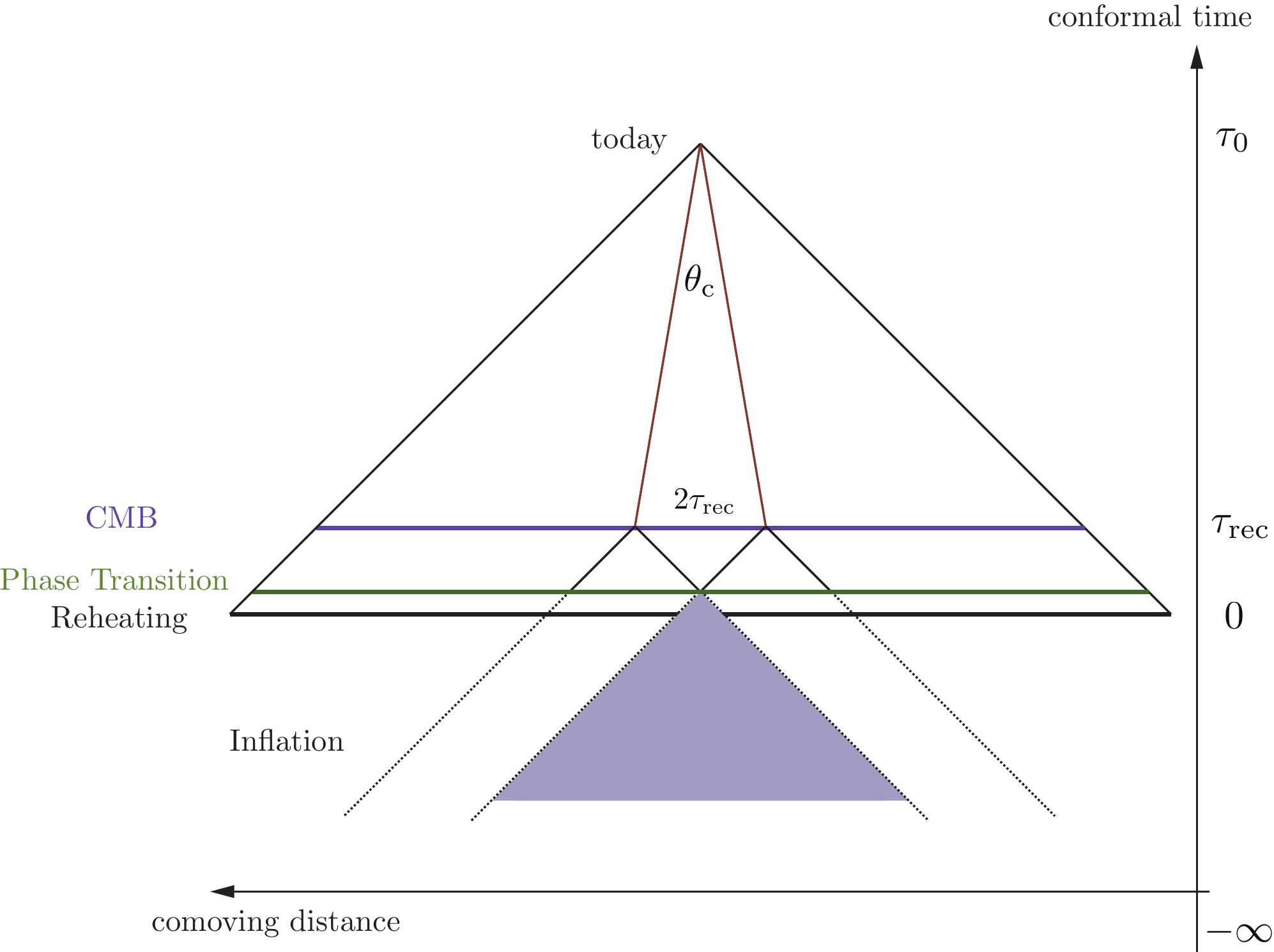}
    \caption{Causal structure of the universe.  Correlations between any local variables at any two spacetime points vanish if their backward light cones fail to intersect on the spacelike hypersurface $\Sigma$ corresponding to the phase transition at $\tau=\tau_{\rm pt}$ \cite{Turok:1996ud}. On the surface of last-scattering at $\tau_{\rm rec}$ this corresponds to angular separations $\theta > \theta_c \approx 2^\circ$. Longer range correlations are established during inflation at negative values of conformal time, $\tau < 0$.}
    \label{fig:causal}
\end{figure}

While both mechanisms are causal, causality imposes a much stronger constraint on tensors from phase transitions or defects than on tensors from inflation (see Figure \ref{fig:causal}).\footnote{In the following we will therefore losely refer to defects and phase transitions as `causal theories' although there is nothing acausal about inflation.}
The causal structure of a (flat) Friedmann-Robertson-Walker (FRW) universe is most conveniently understood by considering its line element in the form
\beq
\d s^2 = a(\tau)^2 \left[ -\d \tau^2 + \d {\bf x}^2\right]\, ,
\eeq
where $\tau = \int \d t/a(t)$ is conformal time which equals the particle horizon at time $t$.
In standard Big Bang cosmology (without inflation) there is a singularity at $\tau=0$.
Inflation extends conformal time to negative values and the time $\tau=0$ becomes the (non-singular) reheating surface. The singularity is pushed to $\tau = - \infty$.
 Recombination takes place at $\tau_{\rm rec}$ and conformal time today is $\tau_0$. 
Let the phase transition occur at a time $\tau_{\rm pt}$ shortly after reheating. The initial perturbation variables induced by the phase transition are defined on a Cauchy surface $\Sigma$ at $\tau_{\rm pt}$.
For causal theories, the unequal time correlator of the source stress energy tensor $T^{\, \rm s}_{\mu \nu}$ satisfies the following constraint
\beq
\langle T^{\, \rm s}_{\mu \nu}(0, \tau)\,T^{\, \rm s}_{\rho \lambda}(r, \tau') \rangle = 0 \quad \qquad \forall \quad r > \tau + \tau'\, .
\eeq
At recombination the particle horizon is $\tau_{\rm hor} \approx \tau_{\rm rec}$, which today corresponds to an angle $\theta_{\rm hor} \approx 1^\circ$ on the sky (see Appendix \ref{sec:horizon}). If we could observe fluctuations at recombination, their angular correlations should therefore satisfy
\beq
C(\theta) = 0 \qquad \forall \quad \theta > \theta_c \equiv 2 \theta_{\rm hor} \approx 2^\circ\, .
\eeq
We recognize that CMB polarization (being generated only by scattering of CMB photons off of free electrons) offers the opportunity to study correlations directly at recombination (this is in contrast to temperature anisotropies which receive contributions after recombination, {\it e.g.}~from the integrated Sachs-Wolfe effect).

\vskip 6pt
In this paper we discuss the superhorizon signature of inflationary tensor modes in the real space $B$-mode correlation function.
This real space treatment has the advantage that causality considerations are applied most directly.
In fact, in harmonic space the difference between the angular power spectra $C_\ell^B$ for inflationary $B$-modes and causally-generated $B$-modes can be more subtle than one would naively anticipate (see Appendix B).
To see this consider the 
scale-dependence of the
power spectra of tensor modes:
inflation, of course, famously produces a nearly scale-invariant spectrum even on superhorizon scales
\beq
\label{equ:PhInf}
P_h(k, \tau) = A_t k^{n_t-3}\, , \qquad n_t \approx 0\, , \qquad \forall \quad k < k_{\rm hor}  = 1/\tau\, .
\eeq
We give causal theories `maximal benefit of the doubt' by allowing them to be tuned to reproduce the inflationary spectrum perfectly on subhorizon scales.
Any difference between inflation and causal theories will then be encoded in a difference in the tensor mode power spectrum for scales near or above the horizon scale.
(In some sense, this is the best causal theories can possibly mimic inflation; in realistic models one often finds a difference even on scales smaller than the horizon.)
On superhorizon scales we asymptotically expect a white noise spectrum for causal theories
\beq
\label{equ:Pc}
P_h^{({\rm c})}(k, \tau) = const.\, , \qquad \forall \quad k < k_{\rm hor} = 1/\tau\, .
\eeq
The angular power spectrum $C_\ell^B$ is a projection of $P_h(k)$ onto the sky.
Ignoring very low multipoles which are dominated by polarization generated by scattering during the epoch of reionization, the low-$\ell$ scaling of the $B$-mode angular power spectrum $C_\ell^B$ is independent of $\ell$ for a scale-invariant input spectrum (\ref{equ:PhInf}); see {\it e.g.}~\cite{Pritchard:2004qp}.
Interestingly, the {\it same} $\ell$-scaling is obtained for the white noise input (\ref{equ:Pc}) from causal theories (see Appendix~B for an explanation of this projection effect).
The difference in the $k$-scalings of inflationary tensors and causal tensors for small $k$ is therefore {\it not} reflected in a difference in the low-$\ell$ scaling of $C_\ell^B$ as one might naively expect.
Any difference between inflationary tensors and causally-generated tensors is encoded in the shape of the angular power spectrum near the peak.
The precise difference will be strongly model-dependent.
To avoid any model-dependent considerations we instead study the unique signature that inflation imprints on the superhorizon $B$-mode correlation in {\it real space}.

\vskip 6pt
To study the causality of $B$-modes in real space we have to address an important subtlety: $B$-modes are defined {\it non-locally} in terms of the Stokes parameters $Q$ and $U$. The causality of $Q$ and $U$ therefore doesn't imply causality of $B$-modes. To avoid this ambiguity we introduce in \S\ref{sec:theory} $\tilde B$-modes as a causal alternative to the conventional $B$-modes.  The $\tilde B$-modes will be defined locally in terms of derivatives of the Stokes parameters $Q$ and $U$ and are therefore manifestly causal by virtue of $Q$ and $U$ being causal.
Like the regular $B$-modes, $\tilde B$-modes are only sourced by tensor (and vector) perturbations and therefore provide an equally good characterization of inflationary gravitational waves.
However, $\tilde B$-modes have the crucial advantage over $B$-modes that causality is studied unambiguously.

\newpage
\section{Causality and Superhorizon Correlations}
\label{sec:theory}

\subsection{Harmonic Expansion}

We recall briefly some basic aspects of the analysis of CMB temperature and polarization anisotropies. More details may be found in the following reviews \cite{reviews}.

\vskip 6pt
The radiation field (temperature $T$ and Stokes parameters $Q$ and $U$) is expanded in spin-0 and spin-2 spherical harmonics \cite{Kamionkowski:1996ks, Zaldarriaga:1996xe}
\begin{eqnarray}
T(\hat n) &=& \sum_{\l m} a_{T,\l m} Y_{\l m}(\hat n) \, ,\\
P_\pm(\hat n) \equiv (Q\pm i U) (\hat n) &=& \sum_{\l m} a_{\pm 2, \l m} \ {}_{\pm 2} Y_{\l m}(\hat n)\, ,
\end{eqnarray}
where the unit vector $\hat n$ denotes the direction on the sky.
The spin-2 polarization field may be transformed into rotationally-invariant spin-0 $E$- and $B$-modes
\begin{eqnarray}
E(\hat n) &=& \sum_{\l m} a_{E,\l m} Y_{\l m}(\hat n) \, ,\\
B(\hat n) &=& \sum_{\l m} a_{B,\l m} Y_{\l m}(\hat n) \, ,
\end{eqnarray}
where
\begin{eqnarray}
a_{E,\l m} &\equiv&  \frac{-1}{2} \left[ a_{2,\l m} + a_{-2,\l m} \right]\, ,\\
a_{B,\l m} &\equiv& \frac{-1}{2i} \left[ a_{2,\l m} - a_{-2,\l m} \right]\, .
\end{eqnarray}
The rotationally-invariant angular power spectra are
\beq
C^{XY}_\l = \frac{1}{2\l + 1} \sum_m \langle a^*_{X,\l m} a_{Y,\l m}\rangle\, ,
\eeq
or
$ \langle a^*_{X,\l m} a_{Y,\l' m'}\rangle = C_\ell^{XY} \delta_{\l \l'} \delta_{m m'} $, where $X, Y = T, E, B$.
The power spectra $C_\ell^{XY}$ are the Legendre transforms of the real space angular correlation functions 
\beq
C^{XY}(\theta) \equiv \langle X(\hat n) Y(\hat n') \rangle = \sum_\ell \frac{2\l +1}{4\pi}\, C_\l^{XY} P_\l(\hat n \cdot \hat n')\, , \qquad \hat n \cdot \hat n' \equiv \cos \theta\, .
\eeq
By statistical isotropy the correlation functions depend only on the angle $\theta$ between the two directions $\hat n$ and $\hat n'$.

\subsection{Causal $\tilde B$-modes}
\label{sec:tildeB}
 

For causal theories (defect models and phase transitions) the real space angular correlation functions of the Stokes parameters vanish outside the particle horizon at recombination, $\theta > \theta_c \equiv 2 \theta_{\rm hor}$.
However, since $E$- and $B$-modes depend {\it non-locally} on the Stokes parameters this causality constraint does not have to hold for $E$- and $B$-mode correlations.
To avoid this ambiguity, it is useful to define the following real space quantities
\begin{eqnarray}
\tilde E(\hat n) &\equiv&  \frac{-1}{2} \left[ \baredth \baredth P_+ + \edth \edth P_- \right] \, , \label{equ:E0}\\
\tilde B(\hat n) &\equiv&  \frac{-1}{2i} \left[ \baredth \baredth P_+ - \edth \edth P_- \right] \, . \label{equ:B0}
\end{eqnarray}
Here, $\edth$ and $\baredth$ are the standard spin-raising and spin-lowering operators, respectively  \cite{Zaldarriaga:1996xe}.
Acting on the spin-2 spherical harmonics they give
\beq
\baredth \baredth {}_{+ 2} Y_{\l m}(\hat n) = n_\ell \, Y_{\ell m}(\hat n) \, , \qquad \edth \edth {}_{- 2} Y_{\l m}(\hat n) = n_\ell \, Y_{\ell m}(\hat n) \, ,
\eeq
where
\beq
\label{equ:nl}
n_\ell \equiv \left[ \frac{(\ell+2)!}{(\ell-2)!}\right]^{1/2}\, .
\eeq
$\tilde E$ and $\tilde B$ therefore have the following harmonic expansions
\begin{eqnarray}
\tilde E(\hat n) &=& \sum_{\l m} a_{\tilde E, \l m} \, Y_{\l m}(\hat n)\, , \label{equ:E}\\
\tilde B(\hat n) &=& \sum_{\l m} a_{\tilde B, \l m} \, Y_{\l m}(\hat n)  \label{equ:B}\, ,
\end{eqnarray}
where
\beq
a_{(\tilde E, \tilde B), \l m} \equiv n_\ell \, a_{(E, B), \l m} \, .
\eeq

Crucially, $\tilde E$ and $\tilde B$ are causal
by virtue of being defined in terms of derivatives of Stokes parameters (which are manifestly causal).
In addition, $\tilde B$-modes are non-zero only for non-zero $B$-modes, so $\tilde B$ is just as good a measure of primordial tensor modes as $B$. However, $\tilde B$ has the advantage over $B$ that causality is checked more directly; ($\tilde B$ has the disadvantage that its noise spectrum is very blue; see \S\ref{sec:signal}).
Finally, we note for completeness, that in the small-scale limit, $\edth \to - (\partial_x + i \partial_y)$, $\baredth \to - (\partial_x - i \partial_y)$, Equations (\ref{equ:E0}) and (\ref{equ:B0}) become
\begin{eqnarray}
\tilde E(\hat n) &\to& -(\partial_x^2 -\partial_y^2) Q - 2 \partial_x \partial_y U \equiv \nabla^2 E\, ,\\
\tilde B(\hat n) &\to& -(\partial_x^2 -\partial_y^2) U + 2 \partial_x \partial_y Q \equiv \nabla^2 B\, ,
\end{eqnarray}
where $\nabla^2$ is the two-dimensional Laplacian defined in the plane of the sky.

\begin{figure}[h!]
    \centering
       \includegraphics[width=0.95\textwidth]{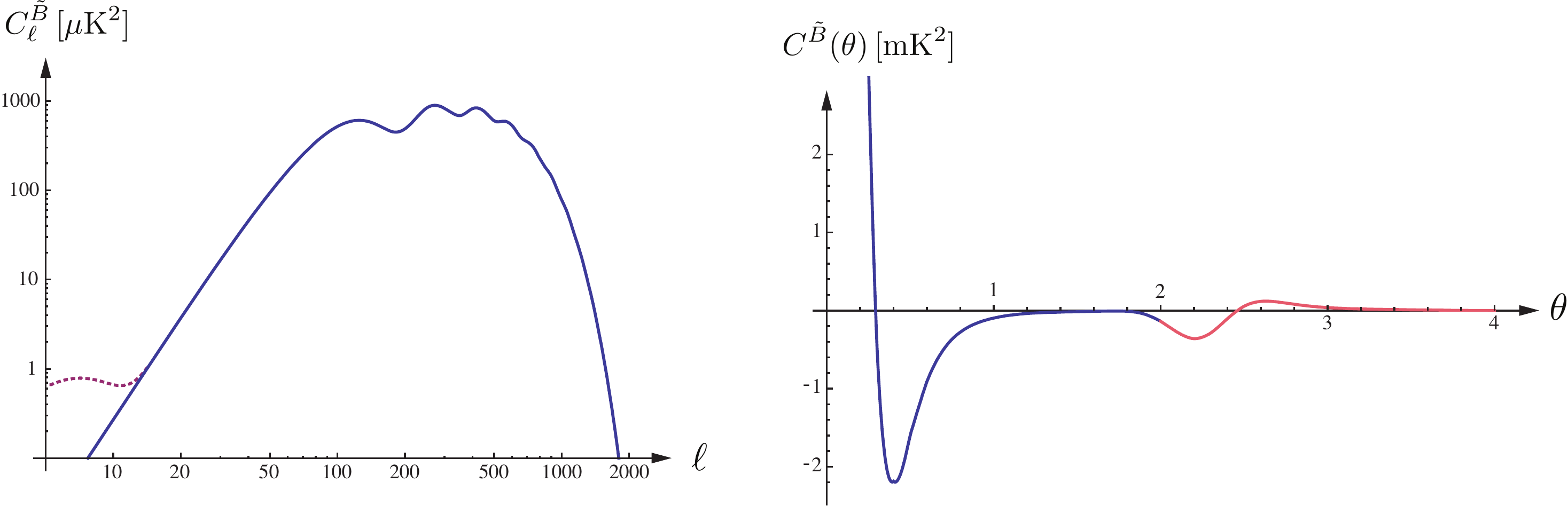}
    \caption{$C^{\tilde B}_\ell \to C^{\tilde B}(\theta)$: transformation from the power spectrum to the real space correlation function. The non-zero $C^{\tilde B}(\theta)$ signal at $\theta \gtrsim 2^\circ$ is a unique signature of inflation.}
    \label{fig:BB}
\end{figure}

For the $\tilde B$-mode correlation function we find
\begin{eqnarray}
C^{ \tilde B}   (\theta)  &=& \sum_\l \frac{2\l +1}{4\pi} \, C_\l^{\tilde B} P_\l(\cos \theta) \,,
\end{eqnarray}
where
\beq
C^{ \tilde B}_\l \equiv  n_\ell^2\, C^{B}_\l   \approx \ell^4 C_\ell^B\, .
\eeq
Similar expressions hold for $C^{T \tilde E}(\theta)$ and $C^{\tilde E}(\theta)$.
The $n_\l^2 \approx \l^4$ conversion factor to go from $C_\l^B$ to $C_\l^{\tilde B}$ has important consequences: small scales (large $\ell$) are weighted more strongly. This has the advantage that superhorizon contributions from the reionization signal at very low-$\ell$ are automatically filtered out; however, it has the disadvantage that the noise spectrum is very blue (see \S\ref{sec:signal}). We have to live with that negative aspect of $\tilde B$-modes, since causality can be studied unambiguously only for $\tilde B$-modes.
One way to see this, is to note that
the real space correlation functions of $\tilde B$-modes and regular $B$-modes are related as follows
\beq
\label{equ:nonlocal}
C^{\tilde B}(\theta) = (\nabla^2 +2)\nabla^2 \, C^B(\theta) \, ;
\eeq
the differential operator $(\nabla^2 + 2) \nabla^2$ becomes $n_\ell^2$ in harmonic $\ell$-space, $(\nabla^2 + 2) \nabla^2 \leftrightarrow n_\ell^2$.
Equation (\ref{equ:nonlocal}) shows that causal $\tilde B$-modes (implied by causal $Q$ and $U$), $C^{\tilde B}(\theta > \theta_c) =0$, do {\it not} imply causal $B$-modes since superhorizon $B$-modes may receive contributions that are in the kernel of $(\nabla^2 + 2) \nabla^2$.
We therefore restrict our discussion of real space correlations and causality to $\tilde B$-modes.


\subsection{Superhorizon Tensor Modes}

For a causal theory we expect
\beq
\label{equ:causal}
C^{\tilde X \tilde Y}(\theta) = 0 \, , \quad {\rm for} \quad  \theta > \theta_c \equiv 2 \theta_{\rm hor}\, ,
\eeq
where $\tilde X, \tilde Y= T,\tilde E, \tilde B$ and $\theta_{\rm hor} \approx 1^\circ$ is the angular separation corresponding to the horizon at recombination.
Note that (\ref{equ:causal}) is a {\it model-independent} constraint that any causal theory has to satisfy.
Temperature fluctuations $T$ and polarization $\tilde E$-modes are sourced dominantly by scalar (density) perturbations.
The litmus test for scalar superhorizon correlations therefore is the $T \tilde E$ cross-correlation on scales corresponding to $\theta > 2^\circ$
\beq
\label{equ:causal2}
C^{T \tilde E}(\theta) \ne 0 \, , \quad {\rm for} \quad  \theta >  \theta_c \approx 2^\circ\, .
\eeq
This is related to but {\it not} the same as the signal discussed in \cite{SZ}.\footnote{Ref.~\cite{SZ} phrased the causality constraints on scalar modes directly in terms of correlations of the Stokes parameters $Q$ and $U$ and their cross-correlations with the temperature fluctuations: for causal theories $C^{QQ}(\theta)$, $C^{UU}(\theta)$ and $C^{TQ}(\theta)$ vanish for $\theta > 2^\circ$.  The proof of the superhorizon nature of adiabatic scalar fluctuations is now often associated with the negative peak of the $TE$ angular power spectrum $C_\ell^{TE}$ for $50 < \ell < 250$ ({\it e.g.}~\cite{Dodelson, WMAP1}), since causal theories (or inflationary models with a significant isocurvature component) tend to predict positive correlations for those multipoles. However, in light of the present discussion superhorizon $TE$ correlations cannot always be unambiguously identified. For $T \tilde E$ correlations this problem is absent.}
Here we propose the analogous 
test for {\it tensor} superhorizon correlations
\beq
\label{equ:causal3}
C^{\tilde B}(\theta) \ne 0 \, , \quad {\rm for} \quad  \theta >  \theta_c \approx 2^\circ\, .
\eeq
Figure \ref{fig:BBx} shows the superhorizon signal from inflation.
Its characteristics are a positive peak at $\theta \gtrsim 2.5^\circ$ and a 
negative peak at $2.0 \lesssim \theta \lesssim 2.5^\circ$.

\begin{figure}[h!]
    \centering
        \includegraphics[width=0.65\textwidth]{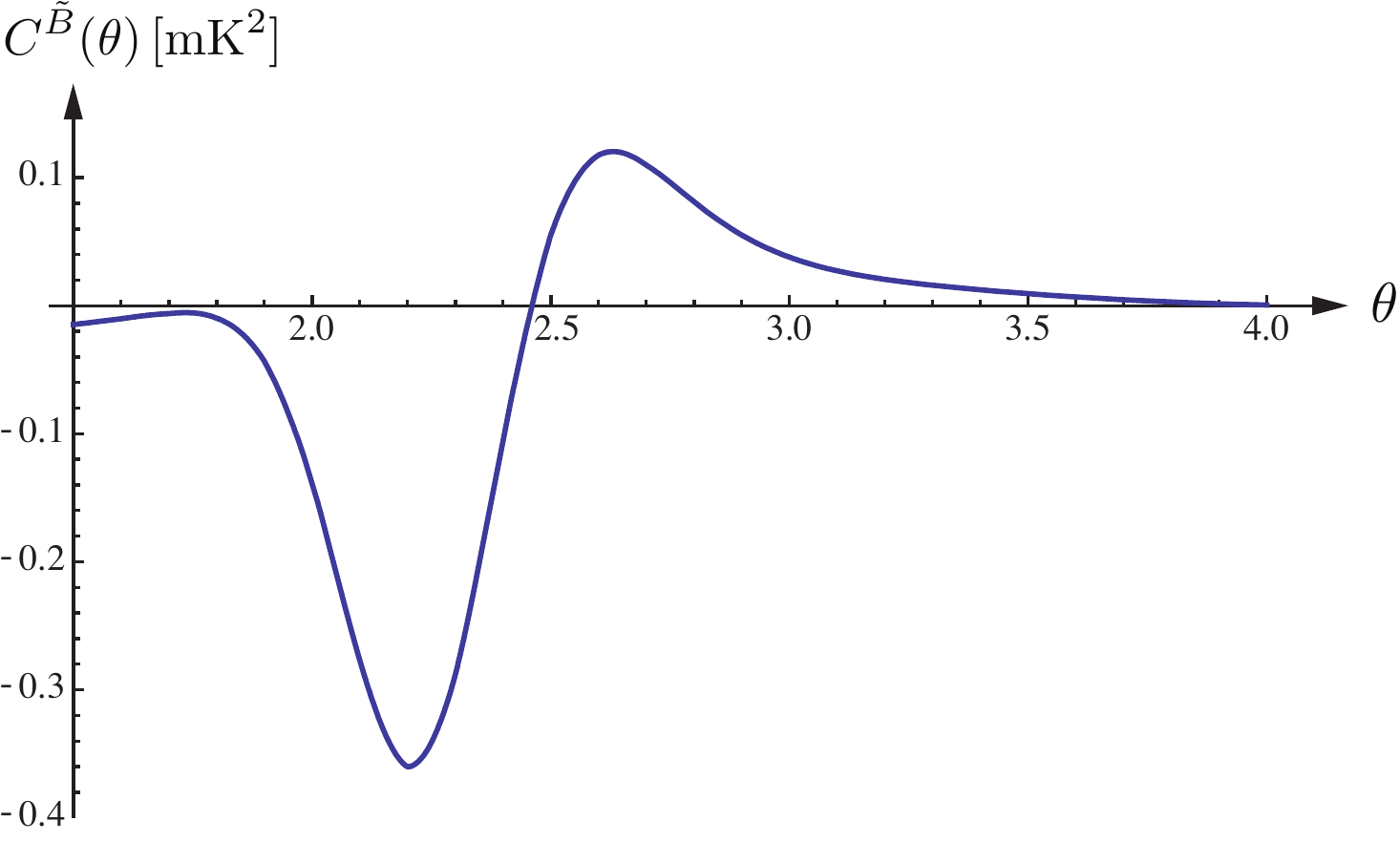}
    \caption{$C^{\tilde B}(\theta) = \langle \tilde B \tilde B \rangle (\theta)$: real space correlation function on superhorizon scales. $C^{\tilde B}(\theta \gtrsim 2^\circ) \ne 0$ is a unique signature of inflationary tensor modes.}
    \label{fig:BBx}
\end{figure}

\newpage
\section{Detectability of the Signal}
\label{sec:signal}

In \S\ref{sec:tildeB} we introduced $\tilde B$-modes as an alternative to the conventional $B$-modes as a tensor mode diagnostic.
A treatment in terms of $\tilde B$-modes is required in order to unambiguously discuss superhorizon correlations in real space.
However, when quantifying the detectability of the inflationary superhorizon signal we have to face the major disadvantage of the $\tilde B$-modes: their noise spectrum is very blue.

\subsection{Noise Spectra and Smoothing}

The noise in the Stokes parameters and in the $B$-modes derived from them is usually assumed to have a {\it white noise} power spectrum (multiplied by the Gaussian window function corresponding to the beam of the experiment)
\beq
N_\ell^B =  \Delta_{P,\rm eff}^{2}\,  e^{\l(\l +1) \sigma_{\rm b}^2}\, ,
\eeq
where $\Delta_{P,\rm eff}$ is the constant noise per multipole and $\sigma_{\rm b} = 0.425\, \theta_{\rm FWHM}$ is the size of the beam. 
Reference sensitivities for representative future CMB polarization experiments are given in Table~1.
We include in the {\it effective} noise spectrum a $B$-mode contribution from the gravitational lensing of $E$-modes.  
Lensing $B$-modes have an approximate white noise spectrum (for $\ell \lesssim 700$) and can therefore be combined with the instrumental noise \cite{Smith:2008an}. 
Without lensing-cleaning ({\it delensing}) the extra noise associated with lensing $B$-modes is $\Delta_{P, \rm lensing} \sim 5\, \mu$K-arcmin. As we will see below, this `lensing noise' would dominate over the inflationary signal, so delensing of the measured $B$-modes is an important constraint on any experiment proposed to measure the superhorizon $\tilde B$-mode signal.
Representative values for the residual noise $\Delta_{P,\rm eff}$ after delensing are shown in Table~1.
The precise values will depend somewhat on the resolution, $\sigma_{\rm b}$, and the instrumental noise, $\Delta_P$, of the experiment. Further details on residual noise after delensing may be found in Ref.~\cite{Smith:2008an, Seljak:2003pn}.
The effective noise should also include residual noise from astrophysical foregrounds like polarized synchrotron and dust radiation from our galaxy \cite{Dunkley:2008am}. We imagine that this is included in $\Delta_{P,\rm eff}$; a more accurate treatment of residual noise from foregrounds and the effects of foregrounds on the delensing is beyond the scope of this paper.

\begin{table}[h]
\begin{center}
\begin{tabular}{ || l  || c | c |  c || }
\hline
{Reference Exp.} & $\Delta_P$ [$\mu$K-${\rm arcmin}$]  & $\Delta_{P,\rm eff}$ [$\mu$K-${\rm arcmin}$]  & $f_{\rm sky}$ \\
 \hline
 \hline
{\sl Planck} 
& 10.00 & -- & 0.7 \\
 \hline
 Balloon & 6.00 & $7.5$ & 0.1\\
   & 3.00 & $4.5$ & 0.1\\
 \hline
{\sl CMBPol}  & 2.00 & $3.0$ & 0.7 \\
& 1.00 & $1.7$  & 0.7\\
 & 0.50 & $1.0$ & 0.7\\
 & 0.25 & $0.7$ & 0.7\\
\hline
no noise & 0 &  0  & 0.7\\
\hline
\end{tabular}
\label{tab:noise}
\caption{Reference sensitivities for future CMB polarization experiments.  
The instrumental noise level is denoted by $\Delta_P$.
Shown are also representative values for the {\it effective} noise levels after {\it delensing} $\Delta_{P,\rm eff}$  \cite{Smith:2008an, Seljak:2003pn}.}
\end{center}
\end{table}

In harmonic space, $\tilde B$-modes and $B$-modes are related by the conversion factor $n_\ell$, see Equation (\ref{equ:nl}).
White noise in the Stokes parameters (or $B$-modes) therefore becomes {\it colored noise} in the $\tilde B$-modes
\beq
N_\l^{\tilde B} = n_\l^2 \, \Delta_{P,\rm eff}^{2}\,  e^{\l(\l +1) \sigma_{\rm b}^2} \approx \ell^4\, \Delta_{P,\rm eff}^2 \, e^{\l(\l +1) \sigma_{\rm b}^2}\, .
\eeq
Because of the $n_\ell^2$--factor in the relation between $\tilde B$-modes and $B$-modes, the
 noise power spectrum for $\tilde B$-modes diverges as $\l^4$. 
 The steep blueness of the noise spectrum means that the noise level is dominated by small-scale noise.
 To regulate the divergence in the noise we therefore smooth both the signal and the noise with a Gaussian smoothing function of width $\sigma_{\rm s}$.  The smoothed spectra are
\beq
\hat C_\ell^{\tilde B} \equiv C_\l^{\tilde B} e^{-\l(\l+1) \sigma_{\rm s}^2}\, , \qquad \hat N_\ell^{\tilde B} \equiv N_\l^{\tilde B} e^{-\l(\l+1) \sigma_{\rm s}^2}\, .
\eeq
Since $\sigma_{\rm s} \gg \sigma_{\rm b}$, we may in practice ignore the finite resolution of the beam.
We also define the total signal as
\beq
\hat {\cal C}_\ell^{\tilde B} \equiv \hat C_\l^{\tilde B} + \hat N_\l^{\tilde B}\, .
\eeq
In Figure \ref{fig:smoothing} we show the dependence of the real space superhorizon correlations on the smoothing scale $\l_{\rm s} \equiv 1/\sigma_{\rm s}$.
Although smoothing decreases the signal it increases the signal-to-noise (see Table~2).
In Figure \ref{fig:BBwNoise} we show the sensitivity to the noise level for a fixed smoothing scale.
We also show the cosmic-variance limit.
Since the signal is much larger than the cosmic-variance, there is in principle no obstacle to measuring it given sufficient instrumental sensitivity.
However, in practice, the strong dependence of the signal-to-noise ratio on $\Delta_{P, \rm eff}$ and $\ell_{\rm s}$ is the major limitation to measuring the superhorizon $\tilde B$-signal from inflation.

\begin{figure}[h!]
    \centering
        \includegraphics[width=0.7\textwidth]{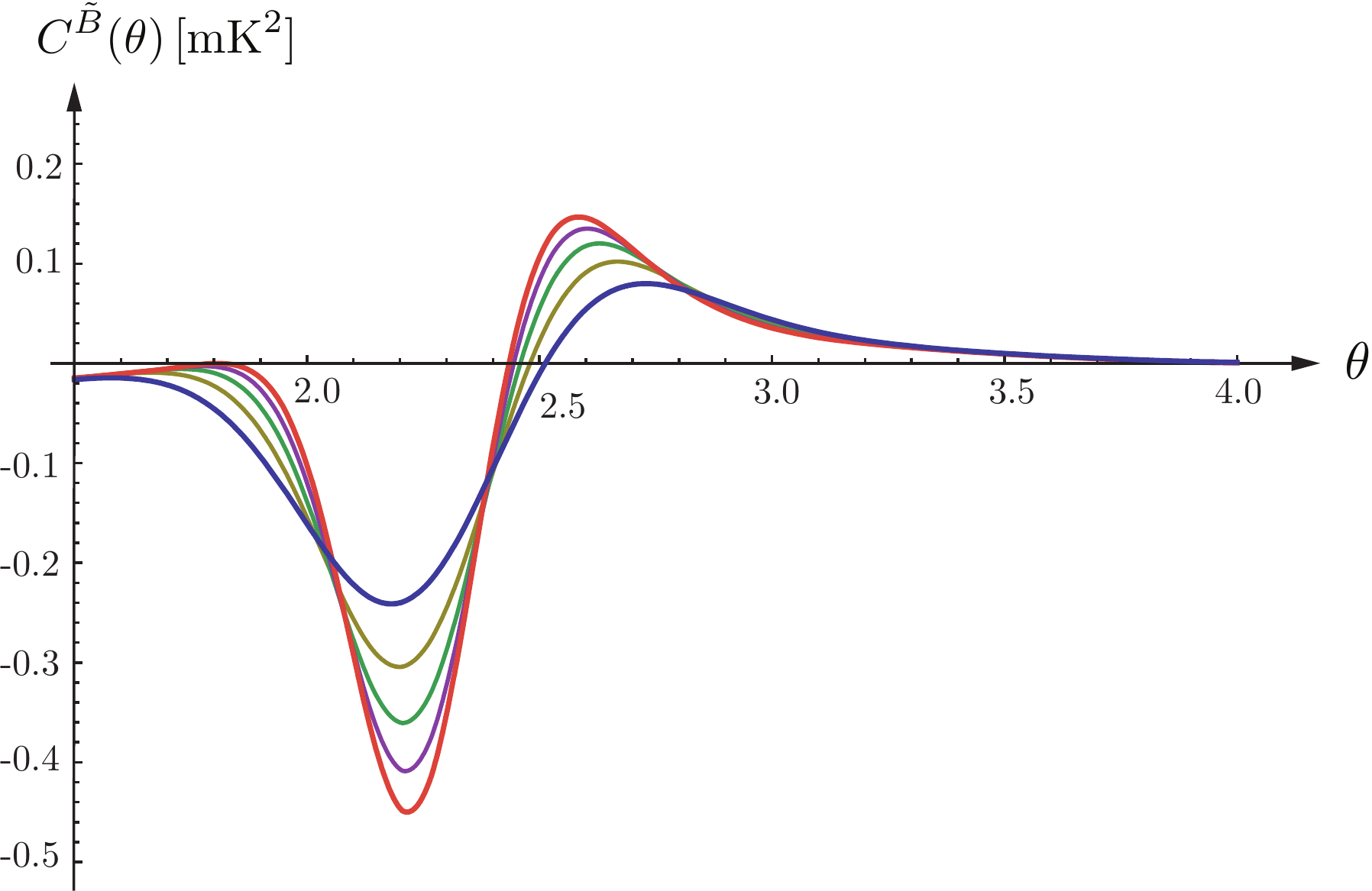}
    \caption{Smoothed signals in real space. The smoothing scales shown range from $\ell_{\rm s}=400$ (blue) to $\ell_{\rm s} = 800$ (red).}
    \label{fig:smoothing}
\end{figure}
\begin{figure}[h!]
    \centering
        \includegraphics[width=0.6\textwidth]{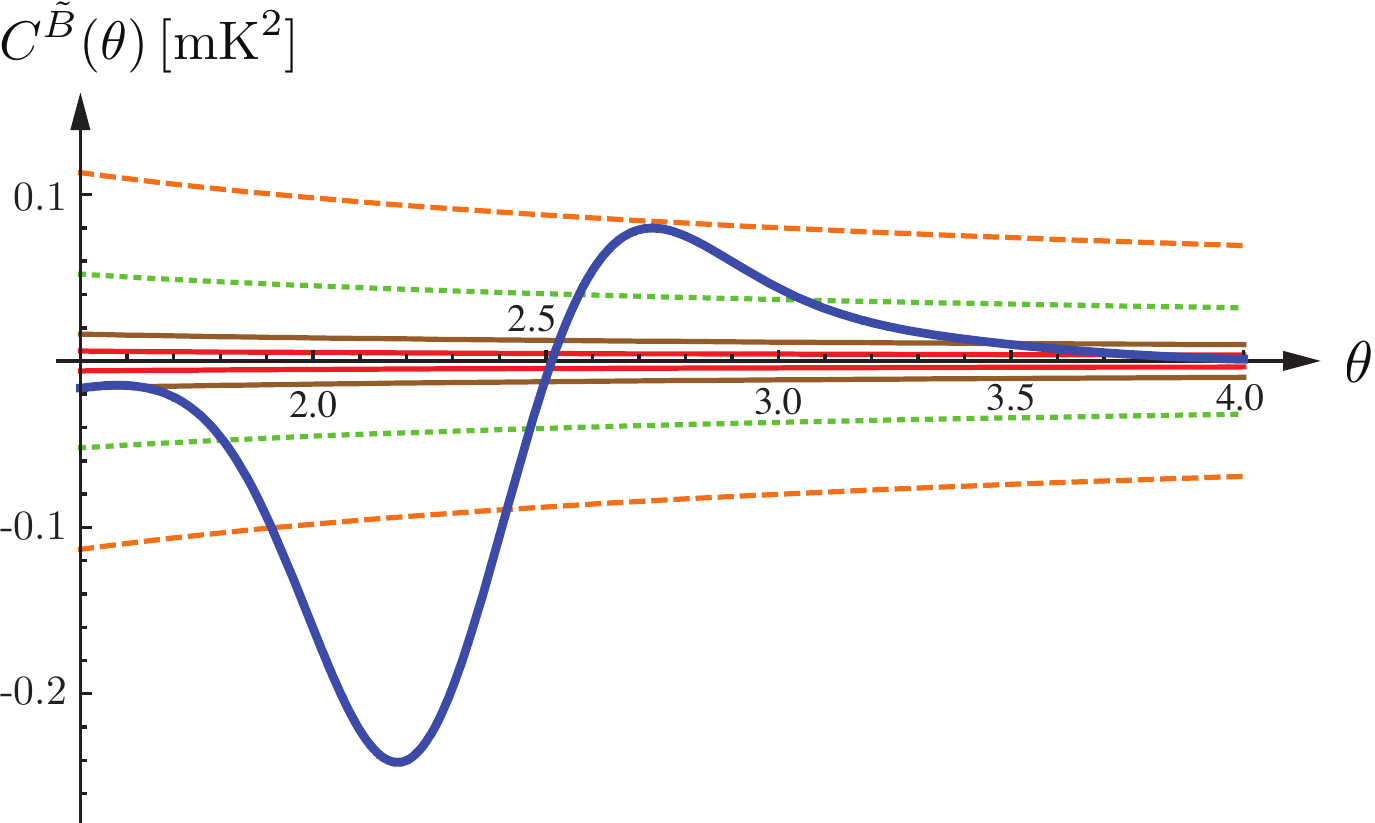}
    \caption{Signal and noise in real space (for a fixed smoothing scale $\ell_{\rm s}=400$).    The  noise levels shown are $\Delta_{P,\rm eff}=0,\ 1,\ 2,\ 3\ \mu$K-arcmin.  Since the signal is much larger than the cosmic-variance, there is in principle no obstacle to measuring it given sufficient instrumental sensitivity. In practice, residual noise after delensing provides a significant constraint.}
    \label{fig:BBwNoise}
\end{figure}

\newpage
\subsection{Superhorizon Signal}

We now quantify the expected signal-to-noise for future experiments.
We will focus specifically on the capabilities of a future CMB polarization satellite ({\sl CMBPol}) \cite{Baumann:2008aj, Baumann:2008aq}.

\vskip 6pt
We bin the superhorizon signal into angular intervals $b \equiv \{\theta_i \pm \frac{1}{2} \Delta \theta\}$, where $\theta_i > \theta_c$ and define a signal vector ${\mathsf S}$ with components $({\mathsf S})_i \equiv \hat C^{\tilde B}(\theta_i)$.
In practice, we use $N_{\rm bin}=10$ bins between $\theta_{\rm min} = 2^\circ$ and $\theta_{\rm max}=4^\circ$.
The different angular bins are of course correlated. The covariance matrix ${\mathsf C}$ has components
\beq
({\mathsf C})_{ij} \equiv \sum_\l \left(\frac{2\l+1}{4\pi}\right)^2 {\rm cov} \left[(\hat {\cal C}_\l^{\tilde B} )^2\right] P_\l(\cos \theta_i) P_\l(\cos \theta_j)\, ,
\eeq
where in harmonic space the covariance matrix is diagonal if the effects of partial sky coverage are ignored
\beq
{\rm cov}\left[\hat {\cal C}_\l^{\tilde B} , \hat {\cal C}_{\l'}^{\tilde B} \right] \equiv \frac{2}{(2\ell +1)f_{\rm sky}}  (\hat {\cal C}_\l^{\tilde B})^2\ \delta_{\ell \ell'}\, .
\eeq
We define the following measure of the signal-to-noise on superhorizon scales
\beq
\label{equ:SN}
\left[\frac{S}{N} \right]^2\equiv {\mathsf S} {\mathsf C}^{-1} {\mathsf S} = \sum_{i,j \, \in\, b} ({\mathsf S})_i ({\mathsf C}^{-1})_{ij} ({\mathsf S})_j\, .
\eeq

\begin{table}[h]
 \label{tab:SN}
\begin{center}
\begin{tabular}{ || l  | c | c | c | c| c|c | c| c|}
\hline
 $\ell_{\rm s}$  &  400 & 500 & 600 & 700 & 800 & 900 & 1000 \\
\hline
 \hline
 $\frac{S}{N}$   & 31.6  & 27.3 & 13.2  & 4.2 & 1.8  & $<1$ & $<1$ \\
 \hline
\end{tabular}
\caption{Signal-to-noise for a {\sl CMBPol}-like experiment, $\Delta_{P, \rm eff} =1\,\mu$K-arcmin $\times \left(\frac{r}{0.1} \right)^{1/2} \times \left(\frac{f_{\rm sky}}{0.7} \right)^{1/2} $, as a function of smoothing scale $\ell_{\rm s}$ for a tensor-to-scalar ratio $r$, sky fraction $f_{\rm sky}$ and number of bins $N_{\rm bin}=10$.}
\end{center}
\end{table}

In Table~2 we evaluate the signal-to-noise for different smoothing scales $\ell_{\rm s}$ and {\it effective} noise levels (instrument noise plus lensing residuals) $\Delta_{P,\rm eff} =1\,\mu$K-arcmin $\times \left(\frac{r}{0.1} \right)^{1/2} \times \left(\frac{f_{\rm sky}}{0.7} \right)^{1/2} $.  For illustration, we assume a fiducial tensor-to-scalar ratio of $r=0.1$ (as expected for $m^2 \phi^2$ inflation and consistent with current upper bounds \cite{Komatsu:2008hk}) and a sky fraction $f_{\rm sky} =0.7$.
Table 2 may then be read as the result for $\Delta_{P,\rm eff} =1\,\mu$K-arcmin. However, we note that
\beq
\left[\frac{S}{N} \right]^2 = f_{\rm sky} \, F(x; \ell_{\rm s})\, , \qquad {\rm where} \quad x \equiv \frac{\Delta_{P,\rm eff} }{\sqrt{r}} \, ,
\eeq
{\it i.e.}~the signal-to-noise ratio only depends on the combination $x\equiv \frac{\Delta_{P,\rm eff} }{\sqrt{r}} $.
Hence, the results in Table~2 can be translated to different values of $r$ by appropriately rescaling the noise level $\Delta_{P,\rm eff}$.
The function $F(x)$ is shown in Figure~6 for different values of $\ell_{\rm s}$.
The signal-to-noise ratio for a no-noise cosmic-variance limited experiment, $S/N \approx 50$, is insensitive to the smoothing.

\begin{figure}[h!]
    \centering
        \includegraphics[width=0.67\textwidth]{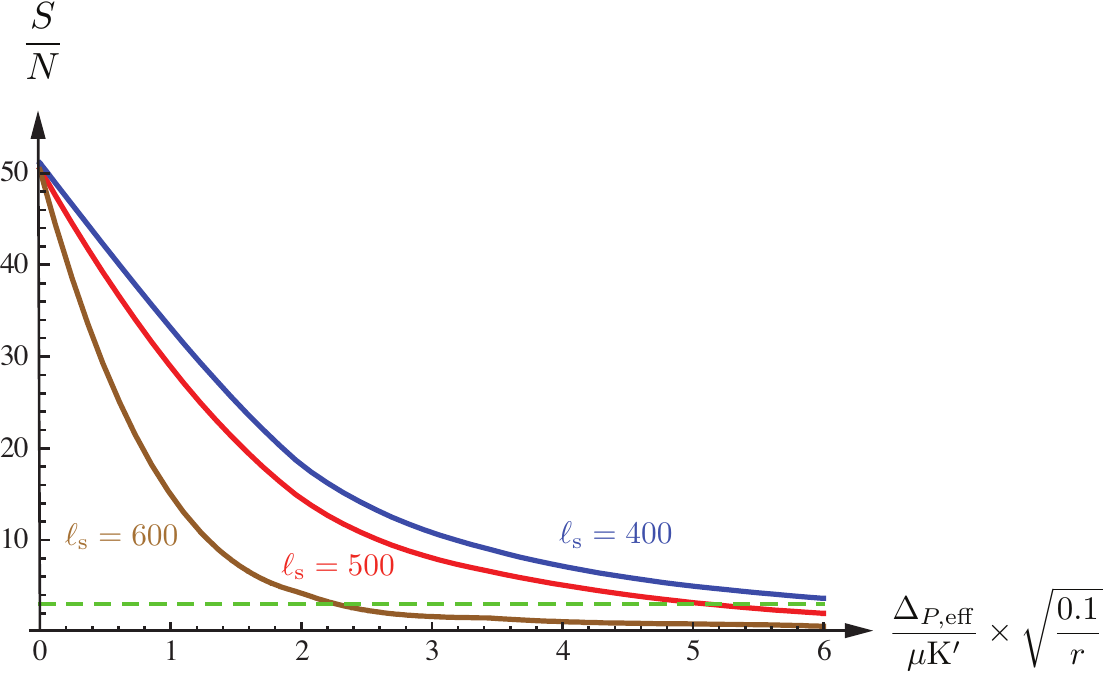}
    \caption{Signal-to-noise as a function of total {\it effective} noise $\Delta_{P, \rm eff}$ (instrument noise plus lensing residuals) and tensor amplitude $r$. Shown are the results for three different smoothing scales. The dashed line indicates $S/N=3$.}
    \label{fig:BB2}
\end{figure}

Finally, in Figure~7 we illustrate the prospects for `signal detection', $S/N \ge 10$ ($\sim$ 10$\sigma$) or $S/N \ge 3$ ($\sim$ 3$\sigma$), as a function of 
$\Delta_{P, \rm eff}$ 
and 
$r$:
\begin{enumerate}
\item[{\sl (a)}] if $r=0.1$ ($0.01$), then the experiment has to allow $\Delta_{P, \rm eff} \lesssim 3.2$ $(1.0)$ $\mu$K-arcmin for $S/N \ge 10$, 
\item[{\sl (b)}] if $r=0.1$ ($0.01$), then the experiment has to allow $\Delta_{P, \rm eff} \lesssim 6.5$ $(2.1)$ $\mu$K-arcmin for $S/N \ge 3$.
\end{enumerate}

We see that even for a relatively large tensor-to-scalar ratio, a detection of the signal 
will require a satellite mission like {\sl CMBPol}.
For smaller $r$, the signal-to-noise decreases roughly as $S/N \propto r$ (for fixed noise $\Delta_{P, \rm eff}$) and becomes proportionately harder to detect.
However, if $r > 0.01$ (the target range for a future {\sl CMBPol} mission \cite{Baumann:2008aq}), then  {\sl CMBPol} would detect the real space superhorizon signal at more than 3$\sigma$ if $\Delta_{P, \rm eff} \approx 2 \mu$K-arcmin (corresponding to instrumental noise $\Delta_P \approx 1\mu$K-arcmin; see Table~1).

\begin{figure}[h!]
    \centering
        \includegraphics[width=0.6\textwidth]{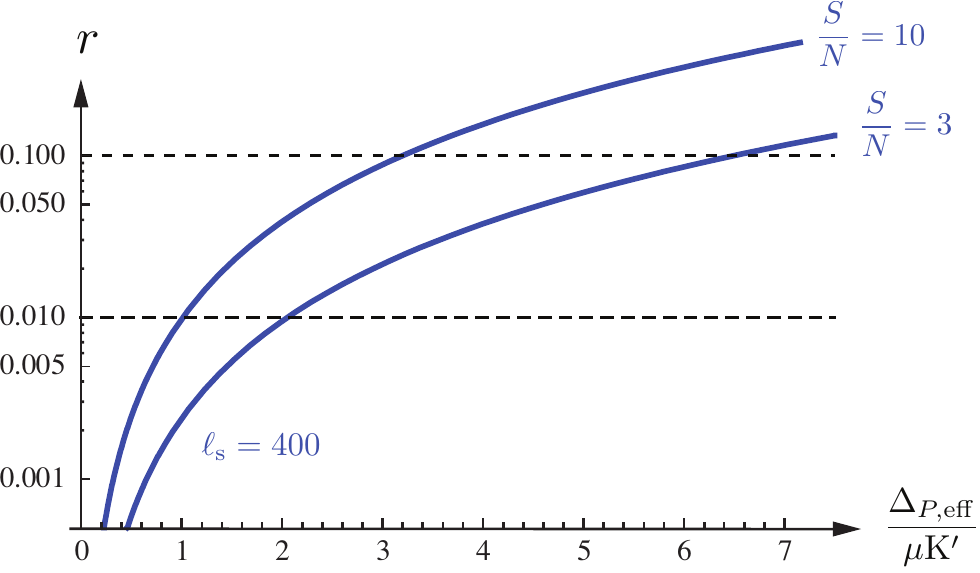}
    \caption{`Signal detection' ($S/N \ge 10$ or $S/N \ge 3$) as a function of total {\it effective} noise $\Delta_{P, \rm eff}$ (instrument noise plus lensing residuals) and tensor amplitude $r$: {\sl (a)} if $r=0.1$ ($0.01$), then the experiment has to allow $\Delta_{P, \rm eff} \lesssim 3.2$ $(1.0)$ $\mu$K-arcmin for $S/N \ge 10$ ($\sim$ 10$\sigma$); {\sl (b)} if $r=0.1$ ($0.01$), then the experiment has to allow $\Delta_{P, \rm eff} \lesssim 6.5$ $(2.1)$ $\mu$K-arcmin for $S/N \ge 3$ ($\sim$ 3$\sigma$).
}
    \label{fig:BB3}
\end{figure}

\newpage
\section{Conclusions}

In this brief note we pointed out that $\tilde B$-mode correlations on angular scales $\theta > 2^\circ$ are an unambiguous signature of inflationary tensor modes.
Since ordinary $B$-modes are defined non-locally in terms of the Stokes parameters and therefore don't have to respect causality, special care had to be taken to define causal $\tilde B$-modes for this analysis. 
The signal can in principle be used to differentiate between $\tilde B$-modes generated by inflation and 
$\tilde B$-modes arising from causally-constrained phase transitions or cosmic strings.
The unambiguous identification of inflationary tensor modes is crucial since it relates directly to the energy scale of inflation. Wrongly associating tensor modes from causal seeds with inflation would imply an incorrect inference of the energy scale of inflation.
In practice, we found that it will be challenging to measure the superhorizon $\tilde B$-mode signal since it requires accurately resolving the recombination peak of the $B$-mode power spectrum.
However, with a future CMB satellite ({\sl CMBPol}) the signal should be detectable if the tensor-to-scalar ratio isn't too small.

We should emphasize that our conclusions are extremely conservative since our analysis assumed that only the superhorizon scales can be used to distinguish the inflationary signal from the signal associated with causal seeds.
In explicit physical models for $B$-modes from cosmic strings or phase transitions one never finds this situation, but there is always a significant difference also in the subhorizon signal, {\it e.g.}~\cite{JonesSmith:2007ne, Seljak:1997ii, Pogosian:2007gi}.
For example, while the inflationary $B$-mode signal peaks on the horizon scale, $B$-modes from causal seeds generically peak on a slightly smaller scale.
For explicit models this difference can be used to distinguish inflationary $B$-modes from $B$-modes associated with causal theories even with {\sl Planck} \cite{Urrestilla:2008jv}.
However, the subhorizon difference between inflationary $B$-modes and causal $B$-modes is 
 model-dependent and therefore not universal.
In this paper we have avoided this model-dependence by focusing exclusively on the unique superhorizon signature of inflationary $\tilde B$-modes.

\subsection*{Acknowledgements}

We thank Amit Yadav and Jaiyul Yoo for useful discussions.
DB thanks Ue-Li Pen for asking questions that inspired this work.

\vskip 40pt

\appendix

\section{Particle Horizon and Causality}
\label{sec:horizon}

In this Appendix we compute the angle subtended by the comoving horizon at recombination.
This is defined as the ratio of the comoving particle horizon at recombination and the comoving angular diameter distance from us (an observer at redshift $z=0$) to recombination ($z \simeq 1090$)
\beq
\theta_{\rm hor} = \frac{d_{\rm hor}}{d_A}\, .
\eeq
A fundamental quantity is the comoving distance between redshifts $z_1$ and $z_2$ 
\beq
\tau_2 - \tau_1 = \int_{z_1}^{z_2} \frac{\d z}{H(z)} \equiv {\cal I}(z_1, z_2)\, .
\eeq
The comoving particle horizon at recombination is
\beq
d_{\rm hor} =  \tau_{\rm rec} - \tau_i \approx {\cal I}(z_{\rm rec}, \infty) \, .
\eeq
In a flat universe, the comoving angular diameter distance from us to
recombination is
\beq
d_A = \tau_0 -\tau_{\rm rec} = {\cal I}(0, z_{\rm rec})\, .
\eeq
The angular scale of the horizon at recombination 
therefore is
\beq
\theta_{\rm hor} \equiv \frac{d_{\rm hor}}{d_A} = \frac{{\cal I}(z_{\rm rec}, \infty)}{{\cal I}(0,z_{\rm rec})}\, .
\eeq 
Using
\beq
H(z)= H_0 \sqrt{ \Omega_m (1+z)^3 + \Omega_\gamma (1+z)^4 + \Omega_\Lambda}\, ,
\eeq
where $\Omega_m =0.27$, $\Omega_\Lambda = 1-\Omega_m$, $\Omega_\gamma = \Omega_m/(1+z_{\rm eq})$ and $z_{\rm eq} = 3400$, we can numerically evaluate the integrals ${\cal I}(0,z_{\rm rec})$ and ${\cal I}(z_{\rm rec}, \infty)$, to find
\beq
\theta_{\rm hor} = 1.16^\circ\, .
\eeq
Causal theories have vanishing correlation functions for 
\beq
\theta > \theta_{\rm c} \equiv 2\theta_{\rm hor} = 2.3^\circ\, .
\eeq

\newpage
\section{CMB Angular Power Spectra}

To gain some intuition for the possible degeneracy between the CMB angular power spectra for inflationary $B$-modes and for causally-generated $B$-modes after reheating we consider simple models for the power spectra of tensor modes:

\begin{itemize}
\item Inflation predicts a nearly scale-invariant spectrum of tensor modes on superhorizon scales
\beq
P_h(k, \tau_i) = A_t k^{n_t -3}\, , \qquad n_t \approx 0\, , \quad \forall \quad k \tau_i < 1\, .
\eeq
The redshifting of gravitational waves when they enter the horizon leads to the following scaling relations for the gravitational wave spectrum at recombination ($\tau_{\rm rec} \equiv 1/k_{\rm rec}$), {\it e.g.}~\cite{Pritchard:2004qp}
\beq
\label{equ:PhR}
\Delta_h(k, \tau_{\rm rec}) \equiv k^3 P_h(k,\tau_{\rm rec}) \propto k^{n_t } \left\{ \begin{array}{ll} 1 & \quad k < k_{\rm rec}  \\
& \\
k^{-4} & \quad k_{\rm rec} < k < k_{\rm eq} \\
& \\
k^{-2} & \quad  k_{\rm eq}  < k
 \end{array} \right.
\eeq
Here, $\tau_{\rm eq} \equiv 1/k_{\rm eq}$ is the time of matter-radiation equality.
Numerically,
\beq
k_{\rm rec} = 50 \, k_0\, , \qquad k_{\rm eq} = 130 \, k_0= 2.7 \, k_{\rm rec}\, .
\eeq
Of course, in reality $\Delta_h$ is a smooth function of $k$. A useful fitting function for a smooth spectrum that asymptotically goes as $k^{n_1}$ for $k< k_\star$ and $k^{-n_2}$ for $k>k_\star$ is
\beq
\Delta_h(k, \tau_\star) =  k^3 P_h(k, \tau_\star) = A\, y^{\frac{n_1-n_2}{2}} \left[ \cosh \left( \frac{\frac{n_1 + n_2}{2} \log y}{\Delta} \right) \right]^{-\Delta}  \, , \qquad y\equiv \frac{k}{k_\star}\, .
\eeq

\item On the other hand, any causal mechanism to produce tensor modes (like global phase transitions or cosmic strings in the early universe) may be tuned to have a scale-invariant spectrum on subhorizon scales at recombination, but by causality we have no reason to believe that it will be scale-invariant on superhorizon scales.
Instead it seems reasonable to expect a {\it Poisson spectrum} (corresponding to uncorrelated white noise) on superhorizon scales
\beq
\label{equ:PhPoisson}
P_{h}^{({\rm c})}(k) = const.  \, , \qquad k < \epsilon \, k_{\rm rec}\, , \quad \epsilon > 1\, .
\eeq
The fudge factor $\epsilon$ parameterizes the peak of the causal tensor mode spectrum (in physical models $\epsilon > 1$, {\it e.g.}~$\epsilon = 3.7$ in \cite{JonesSmith:2007ne}).
The spectrum at recombination therefore is
\beq
\label{equ:PhRc}
\Delta_h^{({\rm c})}(k, \tau_{\rm rec}) \equiv k^3 P_h^{({\rm c})}(k,\tau_{\rm rec}) \propto  \left\{ \begin{array}{ll} k^3 & \quad k < \epsilon\, k_{\rm rec}  \\
& \\
k^{-4} & \quad \epsilon \, k_{\rm rec} < k < \epsilon \,k_{\rm eq} \\
& \\
k^{-2} & \quad \epsilon\, k_{\rm eq} < k
 \end{array} \right.
\eeq

\end{itemize}
Naively, one might think that the difference in the superhorizon power spectra $P_h(k < k_{\rm rec})$ would show up as a difference in the low-$\ell$ CMB power spectra $C_\ell^B$. We now show that this is not the case.

\subsection{Line-of-Sight Formalism}

In the line-of-sight formalism the angular power spectrum may be written as \cite{Seljak:1996is}
\beq
C_\ell^{XY} = (4\pi)^2 \int k^2 \d k \ \underbrace{P_h(k)}_{\rm Spectrum} \ \underbrace{\Delta_{X \ell}(k) \Delta_{Y \ell}(k)}_{\rm Anisotropies}
\eeq
where
\beq
\Delta_{X \ell}(k) \equiv \int_0^{\tau_0} \d \tau \ \underbrace{S_X(k, \tau)}_{\rm Sources} \ \underbrace{P_{X \ell} (k[\tau_0-\tau])}_{\rm Projection}\, .
\eeq
For $B$-mode polarization the source function is 
\begin{eqnarray}
S_B(k, \tau) &\equiv& - g \Psi\, ,
\end{eqnarray}
where $g(\tau)$ 
is the visibility function.
For a Gaussian visibility function 
and assuming slowly varying sources over the surface of last-scattering (of width $\Delta \tau_{\rm rec}$) one furthermore finds
\beq
\Psi \propto \dot h(\tau_{\rm rec}) \Delta \tau_{\rm rec}\, . 
\eeq
The signal is therefore determined by the time-derivative of the gravitational wave amplitude at recombination, $\dot h(\tau_{\rm rec}) \propto k \, h(\tau_{\rm rec})$.
The projection factor for $B$-modes is
\beq
P_{B \ell}(x) \equiv 2 j_\ell'(x) + \frac{4 j_\ell(x)}{x}\, ,
\eeq
where $x\equiv k(\tau_0 - \tau)$.
After approximating the integral over conformal time $\tau$ by the value of its integrand at recombination one finds \cite{Pritchard:2004qp}
\beq
\Delta_{B \ell}(k) \propto P_{B \ell}[k(\tau_0-\tau_{\rm rec})] \dot h(\tau_{\rm rec}) e^{-(k\Delta \tau_{\rm rec})^2/2} \, ,
\eeq
and
\beq
\label{equ:CB}
C^{B}_\ell \propto \int k^4 \d k \ P_h(k, \tau_{\rm rec}) \  P_{B \ell}[k(\tau_0-\tau_{\rm rec})]^2 \ e^{-(k \Delta \tau_{\rm rec})^2}\, .
\eeq



\subsection{Asymptotic Scale-Dependence}

The projection factor $P_{B \ell}(x)$ is peaked at $x \approx \ell$ and for large $k$ scales as $P_{B\ell}^2 \sim k^{-2}$.
This implies that the behavior at last-scattering of the mode with wavenumber $k\approx \ell/(\tau_0- \tau_{\rm rec})$ dominates the contribution to $C_\ell^B$, {\it unless} the integrand grows faster with $k$ than the tail of the projection function (this pathological case will in fact be relevant for $B$-modes from causal theories).

\begin{itemize}
\item For inflation, one can make the following instructive approximation
\begin{eqnarray}
\label{equ:CB2}
C^{B}_\ell &\propto& \left.  k^5 P_h(k, \tau_{\rm rec})\right|_{k\approx \ell/(\tau_0- \tau_{\rm rec})}  \underbrace{\int  \d \ln x  \  P_{B \ell}[x]^2}_{\propto \ \ell^{-2}} \, . 
\label{equ:Capprox}
\end{eqnarray}

From (\ref{equ:PhR}) we may then infer that the polarization power spectrum satisfies the following scalings
\beq
\ell (\ell+1) C_\ell^{B} \propto \left\{ \begin{array}{l l} \ell^2 & \ell < \ell_{\rm rec} \\
& \\
\ell^{-2} & \ell_{\rm rec} < \ell < \ell_{\rm eq} \\
& \\
1 & \ell_{\rm eq} < \ell < \ell_\Delta \\
& \\
\ell^{-4} & \ell_\Delta < \ell
 \end{array}
 \right.
\eeq
\item We have given causal tensors `maximum benefit of the doubt' and assumed that $P_h^{({\rm c})}(k)$ is the same as for inflation for subhorizon scales at recombination, $k > k_{\rm rec}$.
The expected $C_\ell^B$ spectrum is then the same for $\ell > \ell_{\rm rec}$.
Let us compare the spectrum for $\ell \ll \ell_{\rm rec}$.
The approximation (\ref{equ:Capprox}) is now invalid since the integrand grows faster than the decay of the tail of the projection function.
Instead, the integral is 
\beq
\label{equ:CB3}
C^{B}_\ell \propto \int_0^{x_{\rm rec}} x^4 \d x \  P_{B \ell}[x]^2 \ + {\rm small \ correction}\, ,
\eeq
for $\ell < \ell_{\rm rec}$, where $x_{\rm rec} \equiv \frac{k_{\rm rec}}{k_0} \approx 50$. Since for low $\ell$ this converges to a constant independent of $\ell$, we find
\beq
\ell(\ell+1) C_\ell^B \propto \ell^2 \qquad \ell < \ell_{\rm rec}\, ,
\eeq
just as for inflation!
Hence, we conclude that although the power spectra $P_h(k)$, (\ref{equ:PhR}) and (\ref{equ:PhRc}),  are very different for superhorizon scales, $k<k_{\rm rec}$, the projection effects make the $C_\ell^B$ spectra look very similar even at low multipoles, $\ell \ll \ell_{\rm rec}$.
\end{itemize}

Since we assumed that the spectrum on subhorizon scales was tuned to be very close to the spectrum from inflation and we just showed that on superhorizon scales projection effects make the scaling of the $C_\ell^B$ spectra virtually identical, we suspect the causality constraint to be encoded in a difference in the spectra around $\ell \sim \ell_{\rm rec}$.

The precise difference in the $C_\ell^B$ spectra at $\ell \sim \ell_{\rm rec}$ is model-dependent, so instead we focused in this paper on a model-independent signature of superhorizon $B$-modes in the real space correlation function.

\newpage
\vfil

 \begingroup\raggedright\endgroup

 \vfil

\end{document}